\documentclass{article}

\def\be{\begin{equation}}
\def\ee{\end{equation}}
\def\bea{\begin{eqnarray}}
\def\eea{\end{eqnarray}}
\begin{document}
\title{\bf HAMILTONIAN APPROACH TO 2+1 DIMENSIONAL GRAVITY}
\author{L. CANTINI\\
\it Scuola Normale Superiore,  56099 Pisa, Italy\\E-mail:
cantini@cibs.sns.it\\     
P. MENOTTI\\
\it Department of Physics, University of Pisa, 56100 Pisa,
Italy\\E-mail: menotti@df.unipi.it\\   
D. SEMINARA\\
\it Department of Physics, University of Firenze, 50125 Firenze,
Italy\\E-mail: seminara@fi.infn.it}   


\maketitle
\begin{abstract}
It is shown that the reduced particle dynamics
of $2+1$ dimensional gravity in the maximally slicing gauge is of
hamiltonian nature. 
%This is proved directly for the two body problem
%and for the three body problem by using the Garnier equations for
%isomonodromic transformations. 
%For a number of particle greater than
%three the existence of the hamiltonian is shown to be a consequence of
%a conjecture by Polyakov which connects the auxiliary parameters of
%the fuchsian differential equation which solves the $SU(1,1)$ Riemann-
%Hilbert problem, to the Liouville action of the conformal factor which
%describes the space- metric.  
We give the exact diffeomorphism which
transforms the expression of the spinning cone geometry in the Deser,
Jackiw, 't Hooft gauge to the maximally slicing gauge. It is
explicitly shown that the boundary term in the action, written in
hamiltonian form gives the Hamiltonian for the reduced particle
dynamics.  The quantum mechanical translation of the two particle
Hamiltonian is given and the Green function computed.
\end{abstract}
% rise to the logarithm of the Laplace- Beltrami
%operator on a cone whose angular deficit is given by the total energy
%of the system irrespective of the masses of the particles thus proving
%at the quantum level a conjecture by 't Hooft on the two particle
%dynamics.} 
%The quantum mechanical Green's function for the two body
%problem is given.

\section{Introduction}
\label{introd}
Gravity in 2+1 dimensions \cite{DJH} has been the object of vast
interest both at 
the classical and quantum level. 
Several approaches have been
pursued \cite{DJH,hooft,hooft2,wael}. In \cite{BCV,welling} the
maximally slicing gauge, or 
instantaneous York gauge, was introduced. The application of such
a gauge is 
%restricted to universes with spacial topology of genus
%$g<1$ \cite{welling,MS1}; moreover for the sphere topology it can be
%applied only to the static problem \cite{MS2}. Thus the range of
%applicability of such a gauge is 
practically 
restricted to open universes with the topology of the plane; here
however it will prove a very powerful tool.

The approach developed in \cite{BCV,welling} is first order. In
\cite{MS1,MS2} the same gauge was 
exploited in the second order ADM approach; this approach turns out to
be more straightforward than the previous one and being strictly
canonical lends itself to be translated at the quantum
level. 
%Quantization schemes have been proposed in the absence of
%particles in \cite{hooft2,hosoya2,nelsonregge,carlip} and in the
%presence of particles in \cite{hooft2,matschull}.
% 
%
%The present paper is the continuation of two previous  
%papers \cite{MS1,MS2} and goes a lot deeper into the problem.
%
%In sect.2 we 
After giving a concise summary of the results of the previous
papers \cite{MS1,MS2}, we derive in sect.3 generalized
conservation laws starting from the time evolution of the analytic
component of the energy momentum tensor of the Liouville theory which
underlies the conformal factor describing the space metric.
In sect.4 we prove explicitly the hamiltonian nature of the reduced
particle dynamics i.e. the fact that one can give a hamiltonian
description of the time development of the system in terms of the
position and momenta of the particles.
%Thus this is the counterpart of
%the hamiltonian description in the absence of particles for closed
%universes given by Moncrief \cite{moncrief} and Hosoya and Nakao
%\cite{hosoya}.

While for the two particle case the result is elementary, for three 
particles it involves the exploitation of the Garnier equations,
related to the isomonodromic transformations of a fuchsian problem. 
%We
%recall that in \cite{MS1,MS2} it was proved that such Garnier
%equations are an outcome of the ADM dynamical equations of 2+1
%dimensional gravity. 
For more than three particles the proof of the
hamiltonian nature of the reduced equations of motion 
%and the
%derivation of the hamiltonian, 
relies on a conjecture by Polyakov
\cite{ZT1} on the relation between the regularized Liouville action
and the accessory parameters of the $SU(1,1)$ Riemann-Hilbert
problem. Such a conjecture has been proved by Zograf and Takhtajan
\cite{ZT1} for the special cases of parabolic singularities and
elliptic singularities of finite order, but up to now a proof for
general elliptic singularities is absent.

In sect. 5 we give the exact diffeomorphism which relates the conical
metric of Deser, Jackiw and 't Hooft (DJH) in the presence of angular
momentum to its description in the 
maximally slicing gauge.
% as a by-product it gives the exact relation
%between the asymptotic metrics in the DJH and in the maximally
%slicing gauge. These results will be useful in the following to
%understand the boundary terms in the action. We write also the exact
%expression of the Killing vectors in the maximally slicing gauge.

In sect.6 we connect the results of sect.4 with the boundary terms of
the gravitational action; 
%2+1 dimensional gravity coupled to particles
%is an example in which one can compute the Hamiltonian explicitly as a
%boundary term. 
it is explicitly shown that the dynamics is described completely by such 
boundary terms of the action. 

In the last section it is shown that 
%we treat the quantization of the two particle
%problem starting form the classical two particle hamiltonian. The
quantum Hamiltonian for the two body problem 
%turns out to be 
is the logarithm of the
Laplace-Beltrami operator on a cone whose aperture is given by the
total energy of 
the system and is independent of the masses of the two particles. This
provides a complete proof of the conjecture of 't Hooft \cite{hooft3}
about the two particle dynamics.
%{\it i.e.} the equivalence of the 
%relative motion of two particles with that of a test particle on 
%a cone of aperture equal to the total energy. 

%Obviously, the 
%ordering problem is
%always present but the Laplace-Beltrami operator appears to be the
%most natural choice. A very similar structure was found and
%thoroughly examined by Deser and Jackiw \cite{deserjackiw}, when
%treating the quantum 
%problem of a test particle moving on a cone; the main difference is
%that in the present treatment its logarithm rather than the Laplace-
%Beltrami operator appears. 
%
%Given the hamiltonian one can easily compute the Green function; it
%can be written in terms of hypergeometric functions. 
% 
The quantum mechanical problem with more than two particles requires a
more explicit knowledge of the Hamiltonian which is related to the
auxiliary parameters $\beta_B$. The existence of those parameters is
assured by the solvability of the Riemann-Hilbert problem and one can
try to produce a perturbative expansion of them in some
limit situations. Here however the ordering problem is likely to be
more acute.

\section{Hamiltonian approach}

To make the paper relatively self-contained we shall summarize in this
section some results of the papers \cite{MS1,MS2}.
With the usual ADM notation for the metric \cite{ADM}
\begin{equation}
ds^2 = -N^2 dt^2+ g_{ij}(dx^i+ N^i dt)(dx^j+ N^j dt)
\end{equation}
the gravitational action expressed in terms of the
canonical variables is \cite{hawkinghunter}
\begin{eqnarray}\label{gravitationalaction}
&&S_{Grav} = \int dt \int_{\Sigma_t} d^Dx
\left[ \pi^{ij}\dot g_{ij} - N^i H_i - NH\right]-
2\int dt \int_{B_t} d^{(D-1)}x\, r_\alpha
\pi^{\alpha\beta}_{(\sigma_{Bt})} N_\beta\nonumber \\
&&+2\int dt \int_{B_t} d^{(D-1)}x \,\sqrt{\sigma_{Bt}} N 
\left( K_{B_t}+\frac{\eta}{\cosh\eta}{\cal D}_\alpha v^\alpha\right)
\end{eqnarray} 
where for the detailed meaning of the symbols 
%
%
%
%$\sinh\eta=n_\mu u^\mu $ with $n^\mu $ the future
%pointing unit  
%normal to the time slices $\Sigma_t$ and $u^\mu $ the outward pointing
%unit normal to space-like boundary $B$; 
%$B_t= \Sigma_t\cap B$, $\sqrt{\sigma_{Bt}}$ stands for the
%volume form induced by the 
%space metric on $B_t$, $K_{B_t}$ is the extrinsic
%curvature of $B_t$ as a surface embedded in 
%$\Sigma_t$, $v^\alpha\equiv \displaystyle{\frac{1}{\cosh\eta}\left 
%(n^\alpha-\sinh\eta~ u^\alpha\right)}$ and $r_\alpha$ is the versor
%normal to $B_t$ in $\Sigma_t$.    
%The subscript $\sigma_{Bt}$ in $\pi^{\alpha\beta}_{(\sigma_{Bt})}$ is
%a reminder  
%that it has to be considered a tensor density with respect to
%the measure $\sqrt{\sigma_{Bt}}$. 
and the explicit form of $H$ and $H_i$ and of the matter action we refer to 
\cite{MS1}. 
%
%\noindent
%The matter action can be rewritten as 
%\begin{equation} 
%S_m=\int\!d t \sum_n\Big(P_{ni}\, \dot
%q_n^i+N^i(q_n) P_{ni} - N(q_n) \sqrt{P_{ni} P_{nj} g^{ij}(q_n)+
%m_n^2}\Big). 
%\end{equation}
% 
In the $K=0$ gauge and using the complex coordinates $z = x + i y$ the
diffeomorphism constraint is simply solved by 
\begin{equation}\label{pibarzz}
\pi^{\bar z}_{~z} = -\frac{1}{2\pi}\sum_n\frac{P_n}{z-z_n}
\end{equation}
subject to the restriction $\sum_n P_n =0$ \cite{MS2}.
Always for $K=0$ and using the conformal gauge for the space metric
i.e. $g_{ij}dx^idx^j= e^{2\sigma}dz d\bar z$
%\begin{equation}\label{ADMmetric}
%ds^2= -N^2 dt^2+ e^{2\sigma}(dz+ N^z dt)(d\bar z+ N^{\bar z} dt)
%\end{equation}
the hamiltonian constraint takes the form of the
following inhomogeneous Liouville equation  
\begin{equation}
\label{eqsigmatilde2}
2\Delta\tilde\sigma=-e^{-2\tilde\sigma}-4\pi \sum_n \delta^2(z- z_n)(
\mu _n -1)-4\pi\sum_A \delta^2(z- z_A)    
\end{equation}
where $e^{2\sigma} = 2 \pi^{\bar z}_{~z} \pi^{z}_{~\bar z} e^{2\tilde\sigma}$,
%In eq.(\ref{eqsigmatilde2}) $\tilde\sigma$ is defined by
%\begin{equation}
%e^{2\sigma} = 2 \pi^{\bar z}_{~z} \pi^{z}_{~\bar z} e^{2\tilde\sigma},
%\end{equation}
$\mu_n$ are the particle masses divided by $4\pi$, $z_n$ the
particle positions and $z_A$ the positions of the $({\cal N}-2)$ apparent
singularities i.e. of the zeros of eq.(\ref{pibarzz}). 
The Lagrange multipliers $N$ and $N^z$ where expressed in terms of
$\tilde\sigma$ through 
\begin{equation}\label{N}
N = \frac{\partial(-2\tilde\sigma)}{\partial M}
%\end{equation}
%and
%\begin{equation}\label{Nz}
;~~~~N^z =-\frac{2}{\pi^{\bar z}_{\ z}(z)} \partial_z N +g(z),
\end{equation}
with  
\begin{equation}\label{generalg}
g(z) = \sum_B \frac{\partial\beta_B}{\partial M} \frac{1}{z-z_B} 
\frac{{\cal P}(z_B)}{\prod _{C\neq B} (z_B-z_C)} + p_1(z) 
\end{equation}
and  ${\cal P}$ is defined by
\begin{equation} 
-\frac{\pi^{\bar z}_{~z}(z)}{2}= \frac{1}{4\pi}
\sum_n\frac{P_{nz}}{z-z_n}\equiv\frac{\prod_B(z-z_B)}{{\cal P}(z)}. 
\end{equation}
$p_1(z)= c_0(t) + c_1(t) z$ is a first order polynomial. The role
of the first term 
in $g(z)$ is to cancel the poles arising in the first term of
eq.(\ref{N}) due to the zeros of $\pi^{\bar z}_{~z}$ and $\beta_B$
are the accessory parameters of the 
fuchsian differential equation \cite{kra} which underlies the solution
of the Liouville equation (\ref{eqsigmatilde2}).   
The equations for the particle motion are given by \cite{MS1}
\begin{equation}\label{dotz}
\dot z_n = - N^z(z_n) = -g(z_n);~~~~ 
%= -\sum_B
%\frac{\partial\beta_B}{\partial M}\frac{1}{z_n-z_B} 
%\frac{{\cal P}(z_B)}{\prod _{C\neq
%B} (z_B-z_C)} - p_1(z_n)
%\begin{equation}\label{dotP0}
%;~~~~\begin{equation}\label{dotP}
\dot P_{n z} = P_{n a}\frac{\partial N^a}{\partial z} -m_n
\frac{\partial N}{\partial z}.
%\end{equation}  
%\end{equation}
\end{equation}
%$$
%\dot P_{n z} = 4\pi \frac{\partial \beta_n}{\partial M}+P_{nz} ~
%g'(z_n)=
%$$
%\begin{equation}\label{dotP}
%=4\pi \frac{\partial \beta_n}{\partial M} - P_{nz}\sum_B
%\frac{\partial \beta_B}{\partial M} \frac{{\cal P}(z_B)}{(z_n-z_B)^2
%\prod_{C\neq B}(z_B-z_C)}
%+ P_{nz} ~p'_1(z_n).
%\end{equation}
If we want a reference frame which does not rotate at infinity the
linear term in $p_1(z)$ must be chosen so as to cancel in $N^z$ the
term increasing linearly at infinity; such a choice is unique and
given by $-z/(\sum_n P_n z_n)$.

In the simple two particle case the equations of motion in
the relative coordinates $z'_2 = z_2 -z_1$, $P'=P_2 = -P_1$  are 
$\dot z'_2 = 1 /P'_z $ and $\dot P'_z = - \mu/z'_2$ whose 
solution 
\begin{equation}\label{trajectory}
z'_2 = {\rm const}~ [(1-\mu )(t-t_0) - iL/2]^{\frac{1}{1-\mu }}
\end{equation}
agrees with the solution found in \cite{BCV}. 
%A still simpler
%derivation of eq.(\ref{trajectory}) as a ratio of two conservation
%laws, will be given in the next section. 

\section{Virasoro generators and conservation laws}

%In ref.\cite{MS1} the following generalized conservation law (and its
%complex conjugate) for the ${\cal N}$ particle problem was obtained
%\begin{equation}\label{dilconserv}
%\sum_n P_n z_n = (1-\mu)(t-t_0) - iL 
%\end{equation}
%by using the particle equations of motion eq.(\ref{dotz}, \ref{dotP}),
%where $4\pi \mu$ is the total energy of the system and $L$ the
%angular momentum.  
%
%In the two particle case eq.(\ref{dilconserv}) is simply $P' z'_2
%= (1-\mu)(t-t_0) - iL$. As can be easily checked the hamiltonian for
%eqs.(\ref{system}) and 
%their complex conjugates is given by the sum of two conserved
%hamiltonians i.e. $H = h + \bar h$ with $h= \ln(P' {z'}_2^\mu)$. Taking
%the ratio of $P' {z'}_2^\mu = \exp( h ) = {\rm const.}$ with the
%previous equation we obtain the solution eq.(\ref{trajectory}) without
%the need to solve the system (\ref{system}). 
%
%In this section we want to give a treatment of these and analogous
%conservation laws from a more general viewpoint.

In ref.\cite{MS1} the following equation was derived from the ADM
formalism, with regard to the time evolution of the function $Q(z)$
appearing in the fuchsian differential equation underlying $\sigma(z,\bar z)$
\begin{equation}\label{Qequation}
\dot Q(z) = \frac{1}{2}g'''(z) + 2 g'(z) Q(z) +g(z) Q'(z).
\end{equation}
$Q(z)$ can be understood as the analytic component of the energy
momentum tensor of the Liouville theory governing the conformal 
factor $\tilde \sigma$.
%and the above equation
%represents the change of this anomalous energy momentum tensor 
%under a conformal transformation generated by $g(z)$. 
%It was also shown in ref.\cite{MS1} that eq.(\ref{Qequation}) contains
%all the dynamics of the system, i.e. the motion of the particle
%singularities and auxiliary singularities and the change in time of
%the residues at such singularities; it provides also an interpretation
%of 2+1 dimensional gravity \'a la Einstein-Infeld-Hoffmann \cite{EIH}.
We convert
eq.(\ref{Qequation}) into equations for the Laurent series 
coefficients $L_n$ of $Q(z)$ obtaining the following 
%With  
%\begin{equation}
%\frac{1}{2\pi i}\oint z^{n+1} Q(z) dz = L_{n}
%\end{equation}
%we obtain
%\begin{equation}
%L_{-1} = \frac{1}{2}(\sum_n\beta_n +\sum_B \beta_B)
%\end{equation}
%\begin{equation}
%L_{0} = \frac{1}{2}(\sum_n\beta_n z_n +\sum_B \beta_B z_B)+
%        \frac{1}{4}\left[\sum_n (1-\mu_n^2) -3 ({\cal N}-2)\right] 
%\end{equation}
%\begin{equation}
%L_{1} = \frac{1}{2}(\sum_n\beta_n z^2_n +\sum_B \beta_B z^2_B)+
%        \frac{1}{2}(\sum_n (1-\mu_n^2) z_n -3 \sum_B z_B) 
%\end{equation}
%and 
equation of motion
\begin{equation}\label{L-1}
\dot L_{-1} = \frac{c_1}{2}(\sum_n\beta_n +\sum_B \beta_B); 
%\end{equation}
%\begin{equation}\label{L0}
\dot L_{0} = -\frac{c_0}{2}(\sum_n\beta_n +\sum_B \beta_B); 
%\end{equation}
%\begin{equation}\label{L1}
\dot L_{1} = -c_0 L_0 -c_1 L_1.
\end{equation}
%In this paper w
%We shall restrict ourselves to $L_{-1}, L_{0}, L_{1}$. 
The equation for $L_{-1}$ is a simple consistency requirement on the first 
Fuchs relation $\sum_n \beta_n + \sum_B \beta_B =0$.
%We
%recall that $c_0$ and $c_1$ are function of time which specify the 
%translations and roto-dilatations of the reference frame.
%Eq.(\ref{L-1}) is simply a consistency requirement on the first Fuchs
%relation $\sum_n \beta_n + \sum_B \beta_B =0$.
Equation for ${\dot L_0}$ tells us that the value of $L_0$, provided by 
the second Fuchs  relation,
\begin{equation}
\label{L0}
4 L_0 = 2 \sum_n \beta_n z_n + 2 \sum_B \beta_B z_B+
\sum_n (1-\mu_n^2) -3 ({\cal N}-2) = 1 - \mu_\infty^2 = 1 - (1-\mu)^2.
\end{equation}
is constant in time, namely the total mass of the universe is constant.
%This implicitly shows that the total mass $\mu$ is constant in time and
%more importantly, by taking the derivative with respect to $\mu$, we have
%\begin{equation}
%1-\mu = \sum_n\frac{\partial \beta_n}{\partial \mu} z_n
%+\sum_B\frac{\partial \beta_B}{\partial \mu} z_B  
%\end{equation}
%which 
Combining the derivative of (\ref{L0}) with respect to $\mu$ with
 the equations 
of motion provides the generalized 
conservation law, obviously related to the dilatations,
\begin{equation}\label{dilconserv}
\frac{d}{dt}(\sum_nP_n z_n) = 1-\mu.
\end{equation}
%by using the particle equations of motion eq.(\ref{dotz}, \ref{dotP}),
%where $4\pi \mu$ is the total energy of the system and $L$ the
%angular momentum.  
%
In the two particle case eq.(\ref{dilconserv}) is simply $P' z'_2
= (1-\mu)(t-t_0) - iL$. As can be easily checked the Hamiltonian 
in this case  is given by the sum of two conserved
Hamiltonians i.e. $H = h + \bar h$ with $h= \ln(P' {z'}_2^\mu)$. Taking
the ratio of $P' {z'}_2^\mu = \exp( h ) = {\rm const.}$ with the
previous equation we obtain the solution eq.(\ref{trajectory}) without
the need to solve the system (\ref{dotz}). 
Also the equation for $L_{1}$ can be solved; in the non rotating frame  and 
for $c_0(t)=0$ it gives 
\begin{equation}\label{trans}
\sum_n P_n z^2_n = (k_2 t+k_3)[(1-\mu)t-ib]^{\frac{1}{1-\mu}}
\end{equation} 
where $k_2, k_3$ are constants.

\section{Hamiltonian nature of the reduced dynamics}

Starting form the ADM action 
we have reached the particle equations of motion following a canonical 
procedure; thus, we expect equations (\ref{dotz}) to be 
derivable from a Hamiltonian. 
The present section is devoted to the direct proof that such
equations are indeed generated by a Hamiltonian and to the
construction of such Hamiltonian.
Since $\sum_n P_n=0$ it is useful to perform a canonical transformation 
to the variables $z'_n = z_n-z_1,P'_n = P_n, n=2\dots {\cal N}$.
%generated by   
%
%\begin{equation}
%G(z,P')= (z_1+\cdots + z_{\cal N}) P'_1 + (z_2- z_1)P'_2+ \dots +
%(z_{\cal N} - z_1)P'_n,
%\end{equation}
%{\it i.e.}, the change of variables
%\begin{equation}
%z'_1 = z_1+ \dots+ z_{\cal N},\ \  
%z'_2 = z_2 - z_1,\dots\ z'_{\cal N} = z_{\cal N} - z_1,  
%\end{equation}
%
%\begin{equation}
%P'_1 = \frac{P_1 + \dots P_{\cal N}}{\cal N}, 
%\ P'_n = P_n -\frac{P_1 +  \dots + P_{\cal N }}{\cal N},~~~~n>1,
%\end{equation}
%
%which decouples thge relative motion from the global one, described 
%by $z^\prime_1$.
%
%The reduced hamiltonian will be translational invariant,
%i.e. independent of $z'_1$ to be
%consistent with $\sum P_n =0$ and our canonical variables will be
%$z'_2, \dots z'_{\cal N}$ and $P'_2,\dots P'_{\cal N}$.
%
%\noindent
%Using the definition of ${\cal P}(z)$
%\begin{equation}
%\frac{1}{4\pi}
%\sum_n\frac{P_{nz}}{z-z_n}=\frac{\prod_B(z-z_B)}{{\cal P}(z)} 
%\end{equation}
%and 
Using the definition of ${\cal P}$ and the defining property of the locations 
of the apparent singularities $z_B$, $\pi^{\bar z}_z(z_B)=0$, one  finds
%\begin{equation} 
%\frac{\partial z'_B}{\partial P'_n} = \frac{z^\prime_n}
%{4\pi(z'_B-z'_n)z^\prime_B} \frac{{\cal P}(z'_B + z_1)}
%{\prod_{C\neq B}(z'_B -z'_C)},\ \  
%\frac{\partial z'_B}{\partial z'_n} = \frac{-P'_n}{4\pi(z'_B-z'_n)^2}
%\frac{{\cal P}(z'_B + z_1)}{\prod_{C\neq B}(z'_B -z'_C)} 
%\end{equation}
%from which,for $n=2,\dots N$ we have
\begin{equation}\label{dotz1}
\dot z'_n = -\sum_B\frac{\partial\beta_B}{\partial\mu}\frac{\partial
z'_B}{\partial P'_n} -c_1(t) z'_n~~~~
{\rm and}~~~~
\dot P'_n = \frac{\partial\beta_n}{\partial\mu}+
\sum_B\frac{\partial\beta_B}{\partial\mu}\frac{\partial z'_B}{\partial
z'_n} + c_1(t)P'_n. 
\end{equation}
One can get rid of $c_1(t)$ in eq.(\ref{dotz1})
by means of a canonical transformation. 
%of generator is $ G(z',P'')= 
%\sum_na_1(t) z'_n P''_n$ with $\dot a_1(t)=-c_1(t) a_1(t)$. 
%However this holds when  
%$c_1$ is simply a function of $t$. 
%If one wants to write
%eq.(\ref{dotz1}) in the frame which does not rotate at
%infinity, $c_1$ has to be chosen $
%c_1 = -{1/\sum_n P_n z_n}$
%which is not simply a function of $t$. At any rate it is immediately
%seen that if $H$ generates eqs.(\ref{dotz1}) with $c_1=0$ the
%hamiltonian $H+\ln(\sum_n P_n z_n \sum_n \bar P_n \bar z_n)$ generates
%eqs.(\ref{dotz1}) with $c_1$ given by the above expression.
%Thus we shall here examine the case $c_1=0$. 
In the three body case one finds 
that the Hamiltonian is of the form 
\begin{equation}
\label{threehamiltonian}
H(z'_2,z'_3, P'_2,P'_3) = -\int_{z_0}^{z'_A}\frac{\partial
\beta_A}{\partial \mu}(z'_2,z'_3,z''_A) ~d z''_A + f(z'_2, z'_3)
\end{equation} 
where $z'_A$ is a function of $z'_n$ and $P'_n$ through the relation
$\pi^{\bar z}_{~z}(z_B)=0$. Such Hamiltonian provides the equations of motion 
for $\dot z'_n$ and by use of the Garnier equations one proves the existence 
at the local level of the function $f(z'_2, z'_3)$ which provides also the 
equations for $\dot P'_n$.   
\noindent
%We come now to the ${\cal N}$ particle case. 
The natural extension of (\ref{threehamiltonian})  to ${\cal N}$ particles is  
\begin{equation}\label
{HNpart}
H(z'_2,..,z'_{\cal N}, P'_2,..,P_{\cal N})\!=\!\!
-\!\!\int_{\{z_0\}}^{\{z'_B\}}\!\!\sum_B\frac{\partial 
\beta_B}{\partial \mu}(z'_2,..,z'_{\cal N},z''_A,..,\mu)  dz''_B +
f(z'_2,..,z'_{\cal N}).
\end{equation}
In order eq. (\ref{HNpart}) to make sense we need the form 
%that the integral be
%independent of the path in the ${\cal N} - 2$ dimensional space of the
%$z_B$, namely the  form 
$\omega =\sum_A \partial\beta_A/\partial\mu~~ dz_A$ to be exact.
Such a property is a  consequence of a conjecture due to
Polyakov \cite{ZT1} which states 
\begin{equation}\label{Sepsilonmu}
 \sum_n\beta_n dz_n + \sum_B \beta_B dz_B =-\frac{1}{2\pi} d S_\epsilon 
\end{equation}
%that  the accessory parameters in the 
%fuchsian  differential 
%equation which solves the Liouville equation are obtained as
%derivatives of the 
being $S_\epsilon$ regularized Liouville action \cite{takh} 
$$
S_\epsilon [\phi] =\frac{i}{2} \int_{X_\epsilon} (\partial_z\phi 
\partial_{\bar z} \phi +\frac{e^\phi}{2}) dz\wedge d\bar z
-\frac{i}{2}\sum_n(1-\mu_n)\oint_n\phi(\frac{d\bar z}{\bar z -\bar
z_n}- \frac{d z}{ z - z_n})
$$
$$
+\frac{i}{2}\sum_B \oint_B\phi(\frac{d\bar z}{\bar z -\bar
z_B}- \frac{d z}{ z - z_B})
-\frac{i}{2}(\mu-2)\oint_\infty\phi(\frac{d\bar z}{\bar z}- \frac{d
z}{z})
$$
\begin{equation}
-\pi(\sum_n(1-\mu_n)^2  -({\cal N}-2)-(\mu-2)^2)\ln\epsilon^2.
\label{Sepsilon}
\end{equation}
%computed on the solution of the Liouville equation. In
%(\ref{Sepsilon}) $i dz\wedge d\bar z/2 = dx dy$ and $X_\epsilon$ is a
%large disk of radius $1/\epsilon$ from which small 
%disks of radius $\epsilon$ around the particles and apparent
%singularities have been removed. The line integrals are all taken 
%counterclockwise and they impose the correct behavior on $\phi$
%around the singularities and at infinity. Polyakov conjecture then states
%that
%In other words, the accessory parameters $\beta_n$ and $\beta_B$
%which provide $SU(1,1)$ monodromies i.e. a monodromic
%conformal factor, define an exact 1-form. 
Such a conjecture has been
proved by 
Zograf and Tahktajan \cite{ZT1} for fuchsian differential equations
with parabolic singularities; in addition they remark that the proof
can be extended in a straightforward way to the case of elliptic
singularities of finite order.
% We are obviously interested in the
%generic elliptic case 
%including non algebraic singularities (any real $\mu_l$ with
%$0<\mu_l<1$ not necessarily of the form $1/n$). 
The extension of the proof to the general elliptic case is still missing.
%this case seems not as straightforward since the main
%tool of the proof, i.e. the mapping of 
%the upper complex half plane into the punctured Riemann surface
%through a properly discontinuous group, is not available. Nevertheless
%from what follows it appears that such an extension is of great
%relevance for the hamiltonian structure of $2+1$ gravity. 

\noindent
Is is now  straightforward to prove that 
%the hamiltonian 
\begin{equation}\label{Smu}
H = \frac{1}{2\pi}\frac{\partial S_\epsilon}{\partial \mu}\
\end{equation}
is the correct Hamiltonian as
%already provides the correct expression for $\dot
%z'_n$. It is now straightforward to prove that with (\ref{Smu}) also the
%equations for $\dot P_n$ are satisfied. In fact we have
\begin{equation}
-\frac{\partial H}{\partial z'_n} = -\frac{1}{2\pi}\frac{\partial^2
S_\epsilon}{\partial\mu\partial z'_n} -\frac{1}{2\pi}\sum_B\frac{\partial^2
S_\epsilon}{\partial\mu\partial z'_B} \frac{\partial
z'_B}{\partial z'_n} = \frac{\partial
\beta_n}{\partial
\mu}+\sum_B\frac{\partial\beta_B}{\partial\mu}
\frac{\partial z'_B}{\partial z'_n}.  
\end{equation}
%We recall that in the non rotating frame the hamiltonian
%contains an additional contribution, as already observed at the
%beginning of this section. 
The complete form in the non rotating frame is given by
\begin{equation}
H= \ln\left[(\sum_n P_n z_n) (\sum_n \bar P_n \bar z_n)\right] +
\frac{1}{2\pi}\frac{\partial S_\epsilon}{\partial \mu}. 
\end{equation} 
Note that this Hamiltonian, being time-independent,  provides a further
conservation law in the ${\cal N}-$particle problem. For a more complete 
discussion of the relation between the Polyakov conjecture and 
this Hamiltonian we refer the reader to \cite{noi}.

\section{The asymptotic metric}

%In the previous section we constructed the reduced particle
%hamiltonian from the equations of motion. On the other hand one could
%follow a different path, i.e. to recover the hamiltonian as a boundary
%term in the gravitational action. 

In order to understand the boundary terms 
%do so we shall first
we investigate the diffeomorphism which connects the 
metric for a spinning particle in  the DJH gauge 
\begin{equation}\label{DJHmetric}
ds^2 = -(dT+Jd\phi)^2 + dR^2 + \alpha^2 R^2 d\phi^2 = -(dT+Jd\phi)^2 + 
\alpha^2 r_0^2 (\zeta^{\alpha-1})^2 (d\zeta^2 + \zeta^2 d\phi^2),  
\end{equation}
with the metric of the same geometry in the $K=0$ gauge; 
here $R = r_0 \zeta^\alpha$.

%It turns out that such a diffeomorphism can be computed exactly and that 
%will also allow us to compute the expression of the Killing vectors of the 
%spinning cone geometry in our coordinates.
%
%The DJH metric is given by
%
%With the transformation $R = r_0 \zeta^\alpha$ it can be put into
%conformal form 
%\begin{equation}
%ds^2 = 
%\end{equation}
%It is a solution of the 2+1 Einstein's equations
%with a single source located at $r=0$, $\forall t$. 
\noindent
The metric (\ref{DJHmetric}) possesses two Killing vector fields, 
${\partial}/{\partial T}$ and $\partial/\partial \phi$.
%
%For the metric in the maximally slicing gauge we shall use the ADM
%form
%\begin{equation}\label{ADMmetric2}
%ds^2= -N^2 dt^2+ e^{2\sigma}(dz+ N^z dt)(d\bar z+ N^{\bar z} dt).
%\end{equation}
%We shall set with $r=|z|$
%$
%e^{2\sigma}=f^2(r,t);~N^z = z n(r,t);~N^{\bar z}=\bar z \bar n(r,t).
%$
%Such a metric possesses the manifest Killing vector field 
%$\displaystyle{\frac{\partial} {\partial \theta}}$. 
A solution of the Einstein equations which complies to the York
instantaneous gauge and with such Killing vectors is provided by
\begin{equation}\label{piasympt}
\pi^{\bar z}_{~z} = -\frac{1}{2\pi z^2}\sum_n P_n z_n \equiv
\frac{p(t)}{ z},\ \ 
e^{2\sigma} =  \frac{|p(t)|^2}{(z \bar z)n^2}
\frac{16 \alpha^2}{\Lambda(t)^2} \frac{(\frac{z\bar
z}{\Lambda(t)^2})^{-\alpha-1}}{(1- (\frac{z\bar z}{\Lambda(t)^2})^{-\alpha})^2}
\end{equation}
and 
\begin{equation}
\label{N1}
N =\frac{1}{2 \pi\alpha}\left[\frac{1}{2}
\ln \left(\frac{k^2(t) r^{2\alpha}}{\Lambda(t)^{2\alpha}}\right) -1
+\frac{1}{r^{2\alpha}/\Lambda(t)^{2\alpha}-1}
\ln\left(\frac{k^2(t) r^{2\alpha}}{\Lambda(t)^{2\alpha}}\right )
\right] 
\end{equation}
\begin{equation}
\label{Nz1}
N^z = \frac{z}{\pi p(t)}\left[\frac{r^{2\alpha}/\Lambda(t)^{2\alpha}
}{(r^{2\alpha}/\Lambda(t)^{2\alpha}-1)^2}\ln \left(\frac{k^2(t) r^{2\alpha}}{\Lambda(t)^{2\alpha}}\right) 
-\frac{1}{(r^{2\alpha}/\Lambda(t)^{2\alpha}-1)}\right] 
\end{equation}
where $\alpha = 1- \mu  = 1- \frac{M}{4\pi}$ and $r=|z|$.
%Eq.(\ref{piasympt}) reflects the natural asymptotic form of the expression 
%$\displaystyle{
%-\frac{1}{2\pi} \sum_n\frac{P_n}{z-z_n}
%}$  subject to the constraint $\sum_n P_n =0$, while $\sigma$ solves 
%eq. (\ref{eqsigmatilde2}) once we take into account the relation 
%$e^{2\sigma}=2\pi^{\bar z}_{~z}\pi^z_{~\bar z} e^{2\tilde \sigma}$.
%$N$ and $N^z$ are given by eq.(\ref{N},\ref{Nz}). Here we have 
%chosen $g(z)$ so that our reference frame does not rotate at infinity.
While the dependence of $k(t)$ on time is arbitrary,
the canonical equations for $\pi^z_{~\bar z}$ and $\sigma$ fix
the time evolution of $p(t)$ and $\Lambda(t)$ to be
\begin{equation}
p(t) = -\frac{1}{2\pi}[\alpha t -ib],\ \ \ 
\label{Lambda}
\Lambda^2(t) = c_\Lambda [p(t) \bar p(t)]^{\frac{1}{\alpha}}.
\end{equation}
This, in turn, implies that the asymptotic expansion
$
e^{2\sigma} \approx s^2 (z\bar z)^{-\mu}
$
has to be time independent, in agreement with the fact that
 $\ln s^2$ will coincide with the Hamiltonian.
%
%In order to find the diffeomorphism which connects the two metrics,
%one notes that both our metric and that of DJH are independent of 
%the angular variable. This means that the trasformation 
As 
$\partial/\partial \theta$ is mapped into 
$\partial/\partial\phi$ the transformation must have 
the form 
\begin{equation}
\label{diffeo}
R = R(\rho,t);~~~~ T= T(\rho,t);~~~~\phi = \theta +\omega(\rho,t).
\end{equation}
%
%Performing the above change of variables and 
Comparing the independent 
components of the metric, we are led to a set of differential equations
which determines the unknown functions in eq. (\ref{diffeo}) (see \cite{noi} 
for details). At the end one finds  
\begin{equation}
\label{Req} 
R^2 =\frac{1}{\alpha^2} \left[J^2+ \frac{r^{2\alpha}}{4 c_\Lambda(t) \alpha^2}
\left( 1 - c_\Lambda |p(t)|^2 r^{-2\alpha}\right)^2\right] 
\end{equation}
\begin{equation} \label{phieq}
\phi = \theta +\omega \equiv\theta+ \frac{1}{\alpha}\arctan\left[2\pi
\frac{c^{-1}_\Lambda r^{2\alpha} - 
|p(t)|^2 + 2 \alpha^2 J^2}{2 \alpha^2 J t} \right]
\end{equation}
\begin{equation}
\label{Teq}
T = \frac{t}{4\pi} \left[\ln \frac{r^2}{c_\Lambda}
-\frac{1}{\alpha}\ln|p(t)|^2\right] - J \omega +h(t) 
\end{equation} 
with $h(t)$ obeying the relation $\dot h(t)=1/4\pi\alpha [\ln k^2(t)-2\alpha^2 J^2/|p(t)|^2]$.
This shows that two asymptotic solutions with different $k^2(t)$ are
diffeomorphic to the same DJH metric and thus are diffeomorphic to
each other.  This explains why for any choice of $k^2(t)$
eqs.(\ref{piasympt},\ref{N1},\ref{Nz1}) are solutions of Einstein's
equations.  
For large $r$ eqs.(\ref{Req},\ref{phieq},\ref{Teq}) become
\begin{equation} 
R^2 \approx
\frac{r^{2\alpha}}{4c_\Lambda \alpha^4},\ \
\phi \approx \theta + \frac{\pi}{2\alpha},\ \ 
T \approx \frac{t}{4\pi} \ln(\frac{r^2}{c_\Lambda
|p(t)|^{\frac{2}{\alpha}}}) -\frac{\pi J}{2\alpha} +h(t).
\end{equation} 
In the DJH gauge a finite
transformation along the Killing vector
$\partial/\partial T$ is 
simply given by $ T \rightarrow T+ c$ while in the York 
instantaneous gauge it is more complicated. The time-like Killing
vector in the instantaneous York gauge is simply computed and given by 
\begin{equation}\label{killing} 
\frac{(2\pi)^3\alpha (\rho^2+1) |p(t)|^2}{{\cal D}}
\frac{\partial}{\partial t} + \frac{8\pi^2 J \alpha^2}{{\cal
D}}\frac{\partial}{\partial \theta} +
\frac{4\pi \alpha^2 r t}{{\cal D}} \frac{\partial}{\partial r}
\end{equation} 
with
\begin{equation}
{\cal D} = 4\pi^2 |p(t)|^2 (\rho^2+1) [\ln\rho +2\pi\alpha \dot h(t)]
+\alpha^2 t^2(1-\rho^2) + 8 \pi^2 J^2 \alpha^2.  
\end{equation} 
For large $r$ the vector (\ref{killing}) reduces to
\begin{equation}
\frac{4\pi}{\ln(\frac{r^2}{c_\Lambda
|p(t)|^{\frac{2}{\alpha}}})}\left(\frac{\partial}{\partial t} +  
\frac{ \alpha J
c_\Lambda^\alpha}{\pi r^{2\alpha}}\frac{\partial}{\partial \theta} +  
\frac{ \alpha t c_\Lambda^\alpha }{2 \pi^2 r^{2\alpha}}r
\frac{\partial}{\partial r}\right). 
\end{equation} 

\section{The Hamiltonian as a boundary term}

In the $K=0$ conformal gauge we have $\pi^{ij}\dot
g_{ij}\equiv 0$ and thus the action of the particles plus gravity reduces to
\begin{equation}
S  = \int dt (\sum_n P_{ni}\, \dot q_n^i -H_B)
\end{equation}
where $H_B$ is read from eq.(\ref{gravitationalaction}).
%\begin{equation}\label{HB1}
%H_B= -2\int dt \int_{B_t} d^{(D-1)}x \,\sqrt{\sigma_{Bt}} N 
%\left( K_{B_t}+\frac{\eta}{\cosh\eta}{\cal D}_\alpha v^\alpha\right)
%+2\int dt \int_{B_t} d^{(D-1)}x\, r_\alpha
%\pi^{\alpha\beta}_{(\sigma_{Bt})} N_\beta.
%\end{equation} 
%
We want now to extract from $H_B$ the reduced particle Hamiltonian and
compare it to the Hamiltonian $H$ derived directly from the particle
equations of motion.  
It is not difficult to show that at large distances the only surviving 
boundary term is 
%in eq.\ref{gravitationalaction} is the one containing The 
%exterior curvature $K_{Bt}$ of the space boundary.
%   
%The last term in the above equation can be computed as follows: on the
%boundary $x^2+y^2=r_0^2= {\rm const}$ we have
%\begin{equation} 
%r_\alpha \pi^{\alpha\beta}_{(\sigma_{Bt})} N_\beta = -2(\bar z
%\partial_{\bar z} N + z \partial_{z} N) + \bar z g(\bar z) \pi^z_{~\bar
%z} + z g(z) \pi^{\bar z }_{~z} 
%\end{equation}
%whose integral in $d\theta$ between $0$ and $2\pi$ is given by 
%\begin{equation}\label{Nzt}  
%-2 \oint (\bar z \partial_{\bar z} + z \partial_{z}) N d\theta +
%i\oint d\bar z  g(\bar z) \pi^z_{~\bar z} -i \oint dz g(z) \pi^{\bar z
%}_{~z}.
%\end{equation} 
%As for large $|z|$, $N$ behaves like $\ln(z\bar z)/4 \pi$ we see that
%the first term in the above expression goes over to the constant
%$-2$. In the computation of the remaining 
%terms as already noticed in \cite{MS1} the only contribution in $g(z)$
%which survives in the sum is the one arising from the linear term in
%the first order polynomial $p_1(z)$ which in the frame non rotating at
%infinity is given by
%\begin{equation} 
%p_1(z) = c_0 -~\frac{1}{\sum_n P_n z_n} ~z.
%\end{equation} 
%Using this result we find zero for eq.(\ref{Nzt}) i.e. for the last term
%in (\ref{HB1}). 
%Similarly one proves that the contribution of the term ${\cal
%D}_\alpha v^\alpha$ goes to zero like $(r_0^2)^{\mu-1}\ln r_0^2$ for
%$r_0\rightarrow \infty$. 
%Thus we are left with the boundary term is
\begin{equation}\label{HBasint}
H_B  = -2 \int_{B_t} d^{(D-1)}x \,\sqrt{\sigma_{Bt}} N K_{B_t}.
\end{equation} 
By inserting the metric eqs.(\ref{piasympt},\ref{N1},\ref{Nz1}) into
the expression for $K_{Bt}$ 
and $\sigma_{Bt}$ we obtain for the integral
\begin{equation}\label{H_B}
H_B = - 4\pi N r_0 \partial_r[\ln(r e^{\sigma})]
\end{equation} 
and thus for large $r_0$ the boundary term
becomes
\begin{equation}\label{HBr0}
H_B = - r_0\ln r_0^2 (\frac{1}{r_0}+ \partial_r \sigma)=
(\mu-1) \ln r_0^2. 
\end{equation} 
We recall now that the equations of motion are obtained from the
action by keeping the values of the fields fixed at the boundary, or
equivalently  by keeping fixed the intrinsic metric of
the boundary\cite{wald}. In our case the variations should be performed keeping
fixed the fields $N$, $N^a$, and $\sigma$ at the boundary. %Let us
%consider a given value $\mu_0$ of $\mu$. 
We shall perform the
computation for the boundary given by a circle of radius $r_0$ for a
very large value of $r_0$.  If we change the positions of 
particle positions and momenta, $\Lambda$ varies and in order to keep
the value of $\sigma$ fixed at the boundary we must vary $\mu$ as to
satisfy the following equality
\begin{equation}
\ln\{(\sum_n P_n z_n)(\sum_n \bar P_n \bar z_n)\} - \mu \ln r_0^2
+(\mu-1)\ln\Lambda^2 -\ln 16\pi^2 
\equiv -\mu\ln r_0^2 + \ln s^2 = {\rm const.} 
\end{equation} 
Thus
\begin{equation}
0= - \delta \mu \ln r_0^2 +\sum_n( \delta z_n \frac{\partial \ln
s^2}{\partial z_n} + \delta P_n \frac{\partial \ln
s^2}{\partial P_n} + {\rm c.c.} ) + \delta \mu \frac{\partial \ln
s^2}{\partial \mu}
\end{equation} 
i.e. for large $r_0$
\begin{equation}
\delta \mu \approx \frac{1}{\ln r_0^2}\sum_n( \delta z_n \frac{\partial \ln
s^2}{\partial z_n} + \delta P_n \frac{\partial \ln
s^2}{\partial P_n} + {\rm c.c.} ).
\end{equation} 
Substituting % the obtained value of $\mu$ 
into eq.(\ref{HBr0}) %the hamiltonian becomes
we have 
\begin{equation}
\delta H_B = \sum_n( \delta z_n \frac{\partial \ln
s^2}{\partial z_n} + \delta P_n \frac{\partial \ln
s^2}{\partial P_n} + {\rm c.c.} )
\end{equation} 
i.e. apart for a constant $H_B$ equals $ \ln s^2$ 
\begin{equation}\label{Hr}
H_B = \ln s^2 +{\rm const.} = \ln\left[(\sum_n P_n z_n)(\sum_n \bar P_n
\bar z_n)\right] +(\mu-1)\ln\Lambda^2 +{\rm const.}
\end{equation}

%In the two particle case one can check that eq.(\ref{Hr}) coincides
%with the
%hamiltonian derived directly from the equations of motion.
%In fact explicit computation by using the expression of $\Lambda$ in
%terms of hypergeometric functions gives
%\begin{equation}
%\Lambda^2 = |z'_2|^2 \left[ \frac{1}{8(1-\mu)^2
%G(\mu)}\right]^{\frac{1}{1-\mu}}
%\end{equation}
%where
%$$
%G(\mu) = \pi^{-2}\Gamma^4(1-\mu)\sin^2(\pi \mu)\times
%$$ 
%\begin{equation}
%\Delta((\mu+\mu_1+\mu_2)/2)\Delta((\mu-\mu_1+\mu_2)/2)
%\Delta((\mu+\mu_1-\mu_2)/2)\Delta((\mu-\mu_1-\mu_2)/2).  
%\end{equation}
%The boundary term eq.(\ref{HBasint}) depends on the fields on the
%boundary and also on the derivative of the fields directed towards the
%interior i.e. derivative with respect to $r$. By keeping the values of
%the fields fixed on the boundary it provides the hamiltonian, i.e. a
%function of $z_n$ and $P_n$ which through Hamilton's equations give
%rise to the equations of motion. It is not 
\noindent
However the Hamiltonian $H_B$ is not the energy as
usually defined i.e. the value of the boundary term when $(N, N^i)$
take the values of the asymptotic time-like Killing vector. In our
case due to the choice of the $K=0$ gauge which vastly simplifies the
dynamics, the $(N, N^i)$ differ from the timelike asymptotic Killing
vector. The energy of a solution is easily obtained in the DJH gauge,
where one checks from the metric eq.(\ref{DJHmetric}) that with $(N,
N^i)= (1,0,0)$ i.e. the normalized Killing vector, one obtains for $H_B$
the value $4\pi (\mu -1)$ as expected. 

%It is of interest to examine how $\ln\Lambda^2$ behaves 
\noindent
Under a
complex scaling $z' = \alpha z$, $\ln\Lambda^2$
%. It is easily seen  from the Liouville
%equation that if $2\tilde\sigma(z)$ is a solution with singularities
%in $z_n$ and $z_B(z_n, P_n)$ the solution with singularities in
%$\alpha z_n$ and $z_B(\alpha z_n, P_n/\alpha) = \alpha z_B(z_n, P_n)$
%is given by
%\begin{equation}
%2\tilde\sigma(z) = 2\tilde\sigma(\frac{z}{\alpha}) +\ln(\alpha\bar\alpha). 
%\end{equation}
%It implies the following transformation law on $\ln\Lambda$
behaves as follows
\begin{equation}
\ln\Lambda^2(\alpha z_n, \frac{P_n}{\alpha})=\ln\Lambda^2(z_n,P_n)
+\ln(\alpha\bar\alpha)  
\end{equation}
which provides the following Poisson bracket
\begin{equation}
[H,\sum_nP_n z_n] = [\sum_nP_n z_n, (\mu-1) \ln\Lambda^2] = \mu-1 
\end{equation}
and thus we have reached a hamiltonian derivation of the generalized
conservation law
\begin{equation}
\sum_nP_n z_n = (1-\mu)(t-t_0)-ib.
\end{equation}

\noindent
We want now to relate the result eq.(\ref{Hr}) to the results of
sect.4. Let us now consider the derivative of the regularized Liouville action 
with respect to $\mu$; 
%the value of the
%action $S_\epsilon$ on the solution of the Liouville
%equation and let 
%us compute its derivative with respect to $\mu$. A
as we are varying
around a stationary point the only contribution is provided by the
terms in eq.(\ref{Sepsilon}) which depend explicitly on $\mu$ i.e.
\begin{equation}
\frac{\partial S_\epsilon}{\partial \mu} = -\frac{i}{2}
\oint_\infty\phi(\frac{d\bar z}{\bar z}- \frac{d z}{ z}) - 2\pi
(\mu-2)\ln\epsilon^2  
\end{equation}
and as $\phi\equiv -2\tilde\sigma \approx \ln 8(1-\mu)^2 
+(\mu-2)\ln z\bar z-(\mu-1)\ln\Lambda^2$ 
%$ at infinity behaves like
%\begin{equation}
%
%\end{equation}
we have
\begin{equation}
\frac{\partial S_\epsilon}{\partial \mu} = -2\pi \ln8(1-\mu)^2 + 2\pi
(\mu-1)\ln\Lambda^2.
\end{equation}
Thus we can rewrite eq.(\ref{Hr}) as
\begin{equation}
H= \ln\left[(\sum_n P_n z_n) (\sum_n \bar P_n \bar z_n)\right] +
\frac{1}{2\pi}\frac{\partial S_\epsilon}{\partial \mu} + {\rm const}
\end{equation}
in agreement with the result of sect.4 obtained through Polyakov's
conjecture.  
\section{Quantization: the two particle case}
Several quantization schemes for 2+1 dimensional gravity has been proposed
\cite{hosoya2}.

\noindent
%We recall that 
The classical two particle Hamiltonian in the reference
system which does not rotate at infinity given in sect.3 can be written as
%\begin{equation}
%H = \ln(P z \bar P \bar z) + (\mu -1)\ln (z\bar z)=
%\ln(Pz^\mu) + \ln(\bar P\bar z^\mu) = h + \bar h
%\end{equation}
%with $P = P'_2$ and $z = z'_2$.
%$H$ can be rewritten as2
\begin{equation}
H = \ln((x^2+y^2)^\mu ((P_x)^2 + (P_y)^2)).
\end{equation}
Keeping in mind that with our definitions $P$ is the momentum
multiplied by $16\pi G_N/c^3$, applying the correspondence principle 
we have
\begin{equation}
[x,P_x] = [y,P_y] = i l_{P}
\end{equation}
where $l_P = 16 \pi G_n\hbar/c^3$,
all the other commutators equal to zero. $H$ is converted into the
operator
\begin{equation}\label{logbeltrami}
\ln[-(x^2+y^2)^\mu \Delta] +~{\rm constant}.
\end{equation}
i.e. the logarithm is the Laplace-Beltrami
$\Delta_{LB}$ operator on the metric $ds^2=(x^2+y^2)^{-\mu} (dx^2 +
dy^2)$. Following 
an argument similar to the one presented in   
\cite{sorkin} one easily proves that if we start from the 
domain of $\Delta_{LB}$ given by the infinitely  differentiable functions
of compact support $C^\infty_0$ which can also include the origin,
then $\Delta_{LB}$ has a unique 
self-adjoint extension 
%in the Hilbert space of functions square
%integrable on the metric $ds^2=(x^2+y^2)^{-\mu} (dx^2+dy^2)$ 
and 
%as a
%result since $\Delta_{LB}$ is a positive operator, 
$\ln(\Delta_{LB})$ is also self-adjoint (see \cite{noi} for details). 

\noindent
Deser and Jackiw \cite{deserjackiw} have considered the quantum scattering
of a test particle 
on a cone both at the relativistic and non relativistic level. Most of
the techniques developed there can be transferred here:  instead of the 
Hamiltonian $(x^2+y^2)^\mu(p_x^2+p_x^2)$ 
%which appears in their non
%relativistic treatment, 
we have now the Hamiltonian
$\ln[(x^2+y^2)^\mu(p_x^2+p_y^2)]$. The partial wave eigenvalue
differential equation 
\begin{equation}
(r^2)^\mu[-\frac{1}{r} \frac{\partial }{\partial r} r \frac{\partial
}{\partial r}+\frac{n^2}{r^2} ]\phi_n(r) = k^2 \phi_n(r) 
\end{equation}
with $\mu=1-\alpha$ is solved by
$
\phi_n(r) = J_\frac{|n|}{\alpha}(\frac{k}{\alpha}r^\alpha)
$.
A straightforward calculation, using the spectral representation of the above 
operator, leads to the associated Green function, which can be exploited to 
analyze the scattering problem
$$
G(z,z',t) =
\frac{2}{ \alpha\Gamma(\frac{ict}{l_P})r r'} \left (\frac{r^\alpha
+{r'}^\alpha}{2\alpha}\right )^{2 
ict/l_P} 
$$ 
\begin{equation}
\sum_n \frac{e^{in(\phi-\phi')}}{2\pi}
\frac{\Gamma(\frac{|n|}{\alpha}+1 -
\frac{i c t}{l_P})}{\Gamma(\frac{|n|}{\alpha}+1)} 
\rho^{\frac{|n|}{\alpha}+1}{_2F_1}(\frac{|n|}{\alpha}+1 - \frac{ict}{l_P};
\frac{|n|}{\alpha}+\frac{1}{2}; 2 \frac{|n|}{\alpha}+1; 4 \rho)
\end{equation}
where $\rho = r^\alpha {r'}^\alpha/(r^\alpha+{r'}^\alpha)$.

\section*{Acknowledgments}

We are grateful to Marcello Ciafaloni and Stanley Deser for
interesting discussions.

\eject

\end{document}